\documentclass[twocolumn,showpacs,preprintnumbers,amsmath,amssymb]{revtex4}

\usepackage{epsfig}
\usepackage{graphicx}
\usepackage{dcolumn}
\usepackage{bm}

\begin{document}

\preprint{}

\title{Endohedral Impurities in Carbon Nanotubes}

\author{Dennis P. Clougherty}
\email{dpc@physics.uvm.edu}
\affiliation{
Department of Physics\\
University of Vermont\\
Burlington, VT 05405-0125}

\date{September 14, 2002}

\begin{abstract}
A generalization of the Anderson model that includes pseudo-Jahn-Teller impurity coupling is proposed to describe distortions of an endohedral impurity in a carbon nanotube.  Treating the distortion within mean-field theory, spontaneous axial symmetry breaking is found when the vibronic coupling strength $g$ exceeds a critical value $g_c$.  The effective potential in the symmetry-broken state is found to have O(2) symmetry, in agreement with numerical calculations.  For metallic zigzag nanotubes endohedrally-doped with transition metals in the dilute limit, the low-energy properties of the system may display two-channel Kondo behavior; however, strong vibronic coupling is seen to exponentially suppress the Kondo energy scale.
\end{abstract}

\maketitle

Single wall carbon nanotubes \cite{dresselhaus} (SWNTs) are often cited as having the potential to contribute to a number of important technologies, such as electrochemical storage, nanoelectronics, medical imaging, and quantum computing.   In these examples, it is the controlled introduction of impurities, either exohedrally or endohedrally, that are of importance.  Experimentally, as with the fullerenes, SWNTs have been doped by a variety of atoms and molecules.  Recent techniques have led to ``peapod nanotubes,'' where one dimensional arrays of fullerenes  \cite{smith1, smith2}  and metallofullerenes \cite{iijima} have been encapsulated by SWNTs.
 It was subsequently demonstrated \cite{smith3} using scanning tunneling microscopy  (STM) that the endohedral doping of SWNT with fullerenes does indeed modify the local electronic structure of the SWNT.  

For the case of SWNTs endohedrally-doped with atoms or small molecules, Cole and collaborators \cite{cole} have identified two distinct structural phases: the axial phase, where the impurity is confined to the vicinity of the tube's axis of symmetry, and the shell phase, where the impurity is confined to an internal cylindrical surface concentric with the tube.  
For
lithium in a  (5,5) SWNT, the stability of the shell phase has been seen numerically \cite{bernholc}, while the axial phase is the ground state for a system of  H$_2$ molecules at low densities in a (5,5) SWNT \cite{ma}.

Here, the focus is on the development of a theoretical framework for describing a single impurity in a SWNT.  The shell phase can be thought of a resulting from the spontaneous breaking of axial symmetry within mean-field theory.  Such a description is the ferroelastic analog of local magnetic moment formation in a metal within the Anderson model.  The inversion symmetry breaking in metallofullerenes proceeds along similar lines \cite{dpc98}.

It is anticipated  that, in analogy with magnetic impurities in non-magnetic metallic hosts,  the dynamics of the vibronic impurity  in a metallic nanotube will lead to  Kondo-related effects.
In a different context, spin-flip Kondo effects in a carbon nanotube system already have been experimentally explored \cite{nygard}
where the electronic states of a section of a nanotube, analogous to the local electronic states of a quantum dot, play the role of the impurity states. 
Kondo effects for magnetic Co clusters adsorbed on metallic SWNT \cite{lieber} have been discovered using low temperature STM and have been subsequently analyzed theoretically \cite{halperin}.

The vibronic states resulting from endohedral doping and the subsequent Kondo coupling resulting from virtual occupation of those levels give rise to another kind of possible Kondo system.  For strong vibronic coupling, the Kondo temperature, roughly the temperature below which Kondo-related effects are manifested, is exponentially suppressed.

\begin{figure}
\epsfig{file=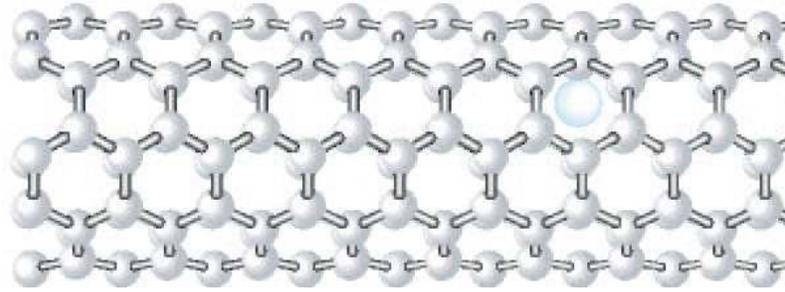,width=6in}
\caption{\label{tube} Sketch of a section of a $(5, 5)$ SWNT with an endohedral atom off-axis (shell phase).}
\end{figure}

There are three classes of electronic band structure for the SWNT \cite{saito}, and they can be characterized by the degeneracy at the Fermi energy.  The first class is semiconducting with no density of states at the Fermi energy. The remaining SWNTs are metallic with the second class having two non-degenerate bands crossing at the Fermi energy away from the zone center; the third class has two doubly degenerate bands crossing at the zone center at the Fermi energy.  The armchair SWNTs $(n,n)$ fall in the second class, and the zigzag SWNTs $(3q,0)$ $q\in Z^+$ fall in the third class, while all others are semiconducting.

For the case where the SWNT has a single non-degenerate band at or near the Fermi energy (the second class and many cases of the first class), the
$a_{1g}$ band electrons may hop to the localized electronic state of the same parity on the impurity site.  This impurity state is taken to be vibronically coupled through the linear pseudo-Jahn-Teller effect to a low-lying state of opposite parity under the transverse displacement $\vec Q$  of the impurity from the axially symmetric position.  
(Actually, this displacement can be regarded as a normal coordinate that includes an accompanying distortion of a nanotube unit cell with the appropriate $e_{2u}$ symmetry.) Vibronic terms that are of ${\cal O}(Q^2)$ and higher are neglected here, as they have been seen to be small in the related endohedral fullerene systems \cite{dpc98}.

Thus, the electronic Hamiltonian is taken as
\begin{equation}
{\cal H}_e=\sum_{k} \epsilon_{k} c^\dagger_{k} c_{k}
+\sum_m E_m d^\dagger_m d_m+\sum_{k} \left(V_{k}c^\dagger_{k} d_{0} + {\rm H.c.}\right)
\end{equation}
where $m=0,\pm 1$ corresponding to the axial angular momentum of the electronic states and the localized one-electron energies have a double degeneracy with $E_1=E_{-1}\equiv E_*$.  
The operators $c^\dagger_{k}$ ($c_{k}$) create (destroy) an electron on the nanotube with wavevector $k$ ($k\ge 0$). The operator  $d^\dagger_{m}$ ($d_{m}$) creates (destroys) an 
electron on the unit cell containing the endohedral dopant.
All energies are relative to the Fermi energy.  The Hamiltonian is diagonal in the spin index.  Thus, spin indices are suppressed.

The discussion here is restricted to the case where the on-site Coulomb repulsion $U$ is sufficiently large that in the limit $V_k=0$ only the singly occupied $m=0$ impurity state is below the Fermi energy $E_f$ (infinite-$U$ model in the local moment regime).  Estimates based on metallofullerene calculations \cite{wastberg} indicate this case to be relevant to alkali-doped nanotubes.

The pseudo-Jahn-Teller contribution to the Hamiltonian must be invariant under symmetry operations of the nanotube. For a nanotube with at least D$_3$ symmetry, the vibronic term for this model has continuous axial symmetry.  The form of the coupling term is determined by symmetry \cite{griffith} up to an overall coupling constant $g$ to be
\begin{equation}
{\cal H}_{JT}=-g\sum_{m=\pm 1}\left(Q_{-m} d^\dagger_m d_0+{\rm H.c.}\right)
\label{arm}
\end{equation}
where $Q_{\pm 1}=Q_x\pm i Q_y$.

Lastly, the form of the elastic energy of the distortion is taken to be
\begin{equation}
{\cal H}_v=\frac{\kappa}{4} \sum_{m=\pm 1} |Q_m|^2
\label{elastic}
\end{equation}

This model is now considered for a classical field $\vec Q$, a mean-field theory in the distortion.  Writing the distortion in polar form
$Q_m=Q\exp(im\phi)$,  the phase $\phi$ is gauged away by transforming $d_m\to d_m\exp(-im\phi)$, leaving only the magnitude $Q$.  
The Hamiltonian is
bilinear in operators and thus is easily diagonalized. The impurity Green's function $G_{00}$  is simply  that of the non-interacting Anderson model with an additional contribution to the self-energy due to the pseudo-Jahn-Teller term
\begin{equation}
G_{00}\equiv \langle 0|G|0\rangle={1\over \omega+i\eta-E_0-\Sigma(\omega)}
\end{equation}
with the self-energy $\Sigma(\omega)$
\begin{equation}
\Sigma(\omega)=\sum_k{|V_k|^2\over \omega+i\eta-\epsilon_k}+{2 g^2 Q^2\over \omega-E_*}
\label{se}
\end{equation}

The energy change due to the vibronic impurity is given by
\begin{equation}
E=\frac{2}{\pi}\int d\omega {\rm Im}\ln G_{00}(\omega)+\frac{\kappa}{2}Q^2
\end{equation}
For systems with coupling strength $g>g_c$, a distorted equilibrium with $Q\ne 0$ is energetically favorable.  All distortion phases $\phi$, however, are degenerate, corresponding to an adiabatic potential with an equipotential circle 
concentric with the symmetry axis.  Such $O(2)$ invariance is broken with the inclusion of higher-order Jahn-Teller couplings.  As numerical calculations \cite{bernholc, ma} point to continuous rotational symmetry of the equilibrium positions of the dopants considered, the linear PJT model proposed is consistent with these results.

In analogy with the Anderson model for magnetic impurities, the broken symmetry is an artifact of the mean-field description.  Phase fluctuations of the distortion stemming from zero point motion of the dopant restore axial symmetry.  

\begin{figure}
\epsfig{file=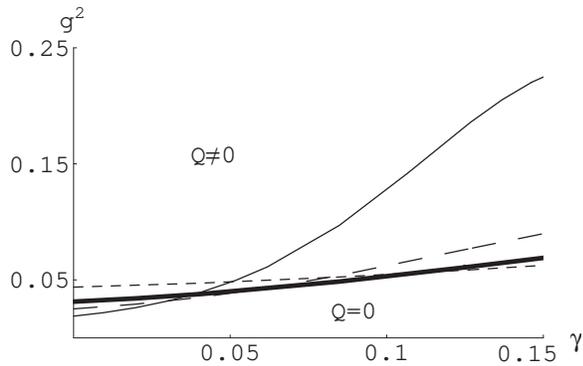,width=3in}
\caption{\label{pd} Phase diagram of vibronic impurity in a SWNT. $g^2$ is measured in units of 
$\kappa t$, $\gamma$ is the resonance width in units of $t$.  Phase boundaries are plotted for $e_* =0.2$, and $e_0=-0.05$ (solid curve), $e_0=-0.1$ (dashed), $e_0=-0.15$ (thick), and $e_0=-0.25$ (dotted).}
\end{figure}

For a linear conduction band with width $2t$, $V_k$ independent of $k$ and $E_0$ near $E_f$, it is straightforward to derive the Stoner-like criterion that
\begin{equation}
{8g_c^2\over\pi\kappa\Gamma} I=1
\label{pb}
\end{equation}
and 
\begin{equation}
I(\gamma, e_*, e_0)=\int_0^1 d\xi {1\over \xi+e_*}{\gamma^2\over(\xi+e_0)^2+\gamma^2}
\label{int}
\end{equation}
where $\gamma$ is the width of the virtual bound state resonance near $E_f$ in units of $t$, $\gamma={\pi |V|^2\rho/ t}$, $e_*=E_*/t$, and $e_0=(E_0-\epsilon)/t$ where $\epsilon$ is the shift in the resonance at $E_f$ due to the hybridization self-energy.
The integral in Eq.~\ref{int} can be expressed in terms of elementary functions, yielding the analytic form for Eq.~\ref{pb} which is plotted in Fig.~\ref{pd}.  

For the case where the SWNT has a doubly degenerate band at or near the Fermi energy (the third class and the remainder of the first class), 
d-states of a transition metal impurity could readily hybridize with the $e_g$ bands of the nanotube, leading to an electronic model of the following form
\begin{eqnarray}
{\cal H}_e&=\sum_{km} \epsilon_{k} c^\dagger_{km} c_{km}
+\sum_{mp} E_p d^\dagger_{mp} d_{mp}\cr
&+\sum_{km} \left(V_{k}c^\dagger_{km} d_{m1} + {\rm H.c.}\right)
\label{2band}
\end{eqnarray}
where $p=\pm 1$ is the parity of the localized orbital doublets.

The impurity states of opposite parity can be coupled through the impurity distortion via the pseudo-Jahn-Teller effect.
\begin{equation}
{\cal H}_{JT}=-g\sum_{m}\left(m Q_{m} d^\dagger_{m1} d_{m -1}+{\rm H.c.}\right)
\label{zigzag}
\end{equation}
As in the case of the Hamiltonian in Eq.~\ref{arm},
the phase of the displacement can be gauged away with the transformation
\begin{equation}
d_{mp}\to d_{mp}{\rm e}^{i\frac{p}{2}(m\phi+(m-1)\frac{\pi}{2})}
\end{equation}
The pseudo-Jahn-Teller contribution to the self-energy in this $(E_u\oplus E_g)\otimes e_u$ model is found to be one-half that of Eq.~\ref{se}. Thus, a qualitatively similar phase diagram is found for the two-band model with $g_c^2\to g_c^2/2$.

For SWNTs of the third class, for $g\le g_c$, orbitally degenerate conduction electrons are coupled to orbital impurity states through virtual fluctuations in impurity occupancy. Orbital states can be labeled by a pseudospin, leading to a possible two-channel Kondo description where the real spin gives rise to the channel indices \cite{cox}; however, the low density of states $\rho$ of metallic zigzag SWNTs tends to lower the Kondo temperature $T_K$ relative to standard metallic hosts.

Under linear vibronic mixing, the ground state double degeneracy of the isolated impurity is preserved.  For finite $U$, provided that fluctuations between the ground state orbital doublet and a doubly occupied, doubly degenerate set of states dominate over fluctuations to the unoccupied singlet, the two-channel Kondo description is still operative; however,  $T_K$  is modified by changes in the hybridization and impurity energy levels.  

This can be quantified in the strong coupling limit $g \gg g_c$.  At low temperatures, fluctuations of the magnitude of the distortion can safely be neglected. Hence, the impurity atom acts as a quantum rotor, displaced from the tube axis by $Q_0$, the mean-field value of the magnitude of the distortion.

The hybridization term can be rewritten as
\begin{equation}
{\cal H}_{\rm a}=\sum_{km}(\tilde V_{k} c^\dagger_{km} |0\rangle\langle m|+{\rm H.c.})
\end{equation}
where $|m\rangle$, $(m=\pm 1)$ are the impurity eigenstates with energy $\xi=E_1+\frac{1}{2}(\delta-\Delta)$, $\delta=E_{-1}-E_1$ and $\Delta^2=\delta^2+4g^2Q_0^2$.
The hybridization is reduced by vibronic mixing in the strong coupling regime to give 
$\tilde V_k=V_k{ \sqrt{\Delta+\delta}\over \sqrt{2\Delta}}$.  
Additionally, with increasing coupling strength, the vibronic energy levels drop, with $\xi\sim -{g^2\over 2\kappa}$ as $g\to\infty$.

Collectively, this implies that the Kondo exchange coupling $J$, as obtained from the generalized Anderson model through a Schrieffer-Wolff transformation, behaves asymptotically as
$J\sim {V^2\kappa\over g^2}$ as $g\to\infty$, leading to $T_K(g)\sim T_K(0)\exp({-g^2/V^2\kappa\rho})$.  Strong vibronic coupling can quickly suppress the temperature scale below which Kondo effects will be manifested. 

In summary, a vibronic generalization of the Anderson model is seen to be a convenient framework for understanding the structural phases of endohedral impurities in SWNTs.  An analogy is noted between the formation of impurity moments in the magnetic Anderson model and the distortion of the endohedral dopant in this vibronic generalization. 

Within mean-field theory, the high symmetry axial phase exists for vibronic coupling $g\le g_c$. For strong coupling, the shell phase is seen to be stable. Beyond mean-field theory, for the models considered where orbital degeneracy becomes vibronic degeneracy, a modification of the two-channel Kondo effect is operative; however, $T_K$ is seen to be exponentially suppressed for strong vibronic coupling.

Support from DARPA under grant no. MDA9720110041 is gratefully acknowledged.

\bibliography{axial}

\begin{thebibliography}{16}
\expandafter\ifx\csname natexlab\endcsname\relax\def\natexlab#1{#1}\fi
\expandafter\ifx\csname bibnamefont\endcsname\relax
  \def\bibnamefont#1{#1}\fi
\expandafter\ifx\csname bibfnamefont\endcsname\relax
  \def\bibfnamefont#1{#1}\fi
\expandafter\ifx\csname citenamefont\endcsname\relax
  \def\citenamefont#1{#1}\fi
\expandafter\ifx\csname url\endcsname\relax
  \def\url#1{\texttt{#1}}\fi
\expandafter\ifx\csname urlprefix\endcsname\relax\def\urlprefix{URL }\fi
\providecommand{\bibinfo}[2]{#2}
\providecommand{\eprint}[2][]{\url{#2}}

\bibitem[{\citenamefont{Dresselhaus et~al.}(1996)\citenamefont{Dresselhaus,
  Dresselhaus, and Eklund}}]{dresselhaus}
\bibinfo{author}{\bibfnamefont{M.~S.} \bibnamefont{Dresselhaus}},
  \bibinfo{author}{\bibfnamefont{G.}~\bibnamefont{Dresselhaus}},
  \bibnamefont{and} \bibinfo{author}{\bibfnamefont{P.~C.}
  \bibnamefont{Eklund}}, \emph{\bibinfo{title}{The Science of Fullerenes and
  Carbon Nanotubes}} (\bibinfo{publisher}{Academic Press},
  \bibinfo{address}{New York}, \bibinfo{year}{1996}).

\bibitem[{\citenamefont{Smith et~al.}(2000)\citenamefont{Smith, Monthioux, and
  Luzzi}}]{smith1}
\bibinfo{author}{\bibfnamefont{B.~W.} \bibnamefont{Smith}},
  \bibinfo{author}{\bibfnamefont{M.}~\bibnamefont{Monthioux}},
  \bibnamefont{and} \bibinfo{author}{\bibfnamefont{D.~E.} \bibnamefont{Luzzi}},
  \bibinfo{journal}{Nature} \textbf{\bibinfo{volume}{396}},
  \bibinfo{pages}{323} (\bibinfo{year}{2000}).

\bibitem[{\citenamefont{Smith and Luzzi}(2000)}]{smith2}
\bibinfo{author}{\bibfnamefont{B.~W.} \bibnamefont{Smith}} \bibnamefont{and}
  \bibinfo{author}{\bibfnamefont{D.~E.} \bibnamefont{Luzzi}},
  \bibinfo{journal}{Chem.\ Phys.\ Lett.} \textbf{\bibinfo{volume}{321}},
  \bibinfo{pages}{169} (\bibinfo{year}{2000}).

\bibitem[{\citenamefont{Hirahara et~al.}(2000)\citenamefont{Hirahara, Suenaga,
  Bandow, Kato, Okazaki, Shinohara, and Iijima}}]{iijima}
\bibinfo{author}{\bibfnamefont{K.}~\bibnamefont{Hirahara}},
  \bibinfo{author}{\bibfnamefont{K.}~\bibnamefont{Suenaga}},
  \bibinfo{author}{\bibfnamefont{S.}~\bibnamefont{Bandow}},
  \bibinfo{author}{\bibfnamefont{H.}~\bibnamefont{Kato}},
  \bibinfo{author}{\bibfnamefont{T.}~\bibnamefont{Okazaki}},
  \bibinfo{author}{\bibfnamefont{H.}~\bibnamefont{Shinohara}},
  \bibnamefont{and} \bibinfo{author}{\bibfnamefont{S.}~\bibnamefont{Iijima}},
  \bibinfo{journal}{Phys.\ Rev.\ Lett.} \textbf{\bibinfo{volume}{85}},
  \bibinfo{pages}{5384} (\bibinfo{year}{2000}).

\bibitem[{\citenamefont{Hornbaker et~al.}(2002)\citenamefont{Hornbaker, Kahng,
  Misra, Smith, Johnson, Mele, Luzzi, and Yazdani}}]{smith3}
\bibinfo{author}{\bibfnamefont{D.~J.} \bibnamefont{Hornbaker}},
  \bibinfo{author}{\bibfnamefont{S.~J.} \bibnamefont{Kahng}},
  \bibinfo{author}{\bibfnamefont{S.}~\bibnamefont{Misra}},
  \bibinfo{author}{\bibfnamefont{B.~W.} \bibnamefont{Smith}},
  \bibinfo{author}{\bibfnamefont{A.~T.} \bibnamefont{Johnson}},
  \bibinfo{author}{\bibfnamefont{E.~J.} \bibnamefont{Mele}},
  \bibinfo{author}{\bibfnamefont{D.~E.} \bibnamefont{Luzzi}}, \bibnamefont{and}
  \bibinfo{author}{\bibfnamefont{A.}~\bibnamefont{Yazdani}},
  \bibinfo{journal}{Science} \textbf{\bibinfo{volume}{295}},
  \bibinfo{pages}{828} (\bibinfo{year}{2002}).

\bibitem[{\citenamefont{Calbi et~al.}(2001)\citenamefont{Calbi, Cole, Gatica,
  Bojan, and Stan}}]{cole}
\bibinfo{author}{\bibfnamefont{M.~M.} \bibnamefont{Calbi}},
  \bibinfo{author}{\bibfnamefont{M.~W.} \bibnamefont{Cole}},
  \bibinfo{author}{\bibfnamefont{S.~M.} \bibnamefont{Gatica}},
  \bibinfo{author}{\bibfnamefont{M.~J.} \bibnamefont{Bojan}}, \bibnamefont{and}
  \bibinfo{author}{\bibfnamefont{G.}~\bibnamefont{Stan}},
  \bibinfo{journal}{Rev.\ Mod.\ Phys.} \textbf{\bibinfo{volume}{73}},
  \bibinfo{pages}{857} (\bibinfo{year}{2001}).

\bibitem[{\citenamefont{Meunier et~al.}(2002)\citenamefont{Meunier, Kephart,
  Roland, and Bernholc}}]{bernholc}
\bibinfo{author}{\bibfnamefont{V.}~\bibnamefont{Meunier}},
  \bibinfo{author}{\bibfnamefont{J.}~\bibnamefont{Kephart}},
  \bibinfo{author}{\bibfnamefont{C.}~\bibnamefont{Roland}}, \bibnamefont{and}
  \bibinfo{author}{\bibfnamefont{J.}~\bibnamefont{Bernholc}},
  \bibinfo{journal}{Phys.\ Rev.\ Lett.} \textbf{\bibinfo{volume}{88}},
  \bibinfo{pages}{75506} (\bibinfo{year}{2002}).

\bibitem[{\citenamefont{Ma et~al.}(2002)\citenamefont{Ma, Xia, Zhao, and
  Ying}}]{ma}
\bibinfo{author}{\bibfnamefont{Y.}~\bibnamefont{Ma}},
  \bibinfo{author}{\bibfnamefont{Y.}~\bibnamefont{Xia}},
  \bibinfo{author}{\bibfnamefont{M.}~\bibnamefont{Zhao}}, \bibnamefont{and}
  \bibinfo{author}{\bibfnamefont{M.}~\bibnamefont{Ying}},
  \bibinfo{journal}{Chem.\ Phys.\ Lett.} \textbf{\bibinfo{volume}{357}},
  \bibinfo{pages}{97} (\bibinfo{year}{2002}).

\bibitem[{\citenamefont{Clougherty and Anderson}(1998)}]{dpc98}
\bibinfo{author}{\bibfnamefont{D.~P.} \bibnamefont{Clougherty}}
  \bibnamefont{and} \bibinfo{author}{\bibfnamefont{F.~G.}
  \bibnamefont{Anderson}}, \bibinfo{journal}{Phys.\ Rev.\ Lett.}
  \textbf{\bibinfo{volume}{80}}, \bibinfo{pages}{3735} (\bibinfo{year}{1998}).

\bibitem[{\citenamefont{Nygard et~al.}(2000)\citenamefont{Nygard, Cobden, and
  Lindelof}}]{nygard}
\bibinfo{author}{\bibfnamefont{J.}~\bibnamefont{Nygard}},
  \bibinfo{author}{\bibfnamefont{D.~H.} \bibnamefont{Cobden}},
  \bibnamefont{and} \bibinfo{author}{\bibfnamefont{P.~E.}
  \bibnamefont{Lindelof}}, \bibinfo{journal}{Nature}
  \textbf{\bibinfo{volume}{408}}, \bibinfo{pages}{342} (\bibinfo{year}{2000}).

\bibitem[{\citenamefont{Odom et~al.}(2000)\citenamefont{Odom, Huang, Cheung,
  and Lieber}}]{lieber}
\bibinfo{author}{\bibfnamefont{T.~A.} \bibnamefont{Odom}},
  \bibinfo{author}{\bibfnamefont{J.-L.} \bibnamefont{Huang}},
  \bibinfo{author}{\bibfnamefont{C.~L.} \bibnamefont{Cheung}},
  \bibnamefont{and} \bibinfo{author}{\bibfnamefont{C.~M.}
  \bibnamefont{Lieber}}, \bibinfo{journal}{Science}
  \textbf{\bibinfo{volume}{290}}, \bibinfo{pages}{1549} (\bibinfo{year}{2000}).

\bibitem[{\citenamefont{Fiete et~al.}(2002)\citenamefont{Fiete, Zarand,
  Halperin, and Oreg}}]{halperin}
\bibinfo{author}{\bibfnamefont{G.~A.} \bibnamefont{Fiete}},
  \bibinfo{author}{\bibfnamefont{G.}~\bibnamefont{Zarand}},
  \bibinfo{author}{\bibfnamefont{B.~I.} \bibnamefont{Halperin}},
  \bibnamefont{and} \bibinfo{author}{\bibfnamefont{Y.}~\bibnamefont{Oreg}},
  \bibinfo{journal}{Phys.\ Rev.\ B} \textbf{\bibinfo{volume}{66}},
  \bibinfo{pages}{024431} (\bibinfo{year}{2002}).

\bibitem[{\citenamefont{Saito et~al.}(1998)\citenamefont{Saito, Dresselhaus,
  and Dresselhaus}}]{saito}
\bibinfo{author}{\bibfnamefont{R.}~\bibnamefont{Saito}},
  \bibinfo{author}{\bibfnamefont{G.}~\bibnamefont{Dresselhaus}},
  \bibnamefont{and} \bibinfo{author}{\bibfnamefont{M.~S.}
  \bibnamefont{Dresselhaus}}, \emph{\bibinfo{title}{The Physical Properties of
  Carbon Nanotubes}} (\bibinfo{publisher}{Imperial College Press},
  \bibinfo{address}{London}, \bibinfo{year}{1998}).

\bibitem[{\citenamefont{W{\"a}stberg and Rosen}(1988)}]{wastberg}
\bibinfo{author}{\bibfnamefont{B.}~\bibnamefont{W{\"a}stberg}}
  \bibnamefont{and} \bibinfo{author}{\bibfnamefont{A.}~\bibnamefont{Rosen}},
  \bibinfo{journal}{Phys.\ Scr.} \textbf{\bibinfo{volume}{44}},
  \bibinfo{pages}{276} (\bibinfo{year}{1988}).

\bibitem[{\citenamefont{Griffith}(1962)}]{griffith}
\bibinfo{author}{\bibfnamefont{J.~S.} \bibnamefont{Griffith}},
  \emph{\bibinfo{title}{The Irreducible Tensor Method for Molecular Symmetry
  Groups}} (\bibinfo{publisher}{Prentice Hall}, \bibinfo{address}{New Jersey},
  \bibinfo{year}{1962}).

\bibitem[{\citenamefont{Cox}(1987)}]{cox}
\bibinfo{author}{\bibfnamefont{D.~L.} \bibnamefont{Cox}},
  \bibinfo{journal}{Phys.\ Rev.\ Lett.} \textbf{\bibinfo{volume}{59}},
  \bibinfo{pages}{1240} (\bibinfo{year}{1987}).

\end{thebibliography}
\end{document}